\documentclass[pra,amsmath,twocolumn,showpacs]{revtex4-1}

\usepackage{graphicx}
\usepackage{amsmath, amsthm, amssymb}
\usepackage{bm}
\usepackage{amssymb}
\usepackage{mathtools}
\usepackage[dvipsnames]{xcolor}
\usepackage{epigraph}
\usepackage{csquotes}
\usepackage{enumitem}
\usepackage{physics}
\usepackage{mathrsfs}
\usepackage{dsfont}
\usepackage{float}
\usepackage{changepage}
\usepackage[toc,page]{appendix}
\usepackage[hidelinks]{hyperref}
\hypersetup{
	colorlinks=true,
	linkcolor=Blue,
	citecolor=Blue, %PineGreen, OliveGreen
	filecolor=Black,      
	urlcolor=Blue, %BlueViolet, %RoyalPurple
}
\begin{document}
	
%%%%%%%%%%%%%%%%%%%%%%%%%%%%%%%%%%%%%%%%%%

\def\la{\langle}
\def\u{\hat U}
\def\A{\mathcal {A}^s}
\def\a{ a}
\def\B{\hat B}
\def\C{\hat C}
\def\F{\tilde F}
\def\zz{\tilde z}
\def\dd{\Delta f}
\def\d{\Delta f}
\def\Om{\Omega}
\def\up{\uparrow}
\def\do{\downarrow}
\def\ep{\epsilon}
\def\+{\uparrow}
\def\-{\downarrow}
\def\r{'\ref'}
\def\I{{\text {Im}}}
\def\R{\text{Re}}
\def\fb{\overline F}
\def\pia{\hat \pi_A}
\def\pin{\hat \pi}
\def\wb{\overline W}
\def\nl{\newline}
\def\h{\hat H}
\def\lm{\lambda}
\def\lmu{\underline\lambda}
\def\q{\quad}
\def\t{\tau}
\def\om{\omega}
\def\rr{\tilde \rho}
\def\p{\hat  \pi}
\def\s{\mathcal{S}}
\def\rrr{\color{red}}
\def\g{\color{blue}}
\def\yy{\colorbox{yellow}}
\def\n{\\ \nonumber}
\def\ra{\rangle}
\def\Ep{{\mathcal{E}}}
\def\Ep{{\mathcal{E}}}
\def\E{{\epsilon}}
\def\h{\hat{H}}
\def\ha{\hat{H}_A}
\def\ua{\hat U_A}
\def\e{\enquote}
\def\f{\vec {f}}
\def\aa{\vec {a}}
\def\b{\vec {b}}
\def\ep{\epsilon}
\def\ve{\varepsilon}
\def\1{\mathds{1}}
\def\AA{\mathcal {A}}
\def\bb{\vec {b}}
\def\th{\thinspace\thinspace}
%=================================================================
% Please use the following mathematics environments: Theorem, Lemma, Corollary, Proposition, Characterization, Property, Problem, Example, ExamplesandDefinitions, Hypothesis, Remark, Definition, Notation, Assumption
%% For proofs, please use the proof environment (the amsthm package is loaded by the MDPI class).

%=================================================================
% Full title of the paper (Capitalized)
\title{Quantum weak values and the \e{which way?} question}

% MDPI internal command: Title for citation in the left column

%\date\today
\author {A. Uranga$^{a,b}$} 
%\email {dgsokol15@gmail.com}
\author {E. Akhmatskaya$^{a,c}$}
\author {D. Sokolovski$^{b,c,d}$}
%\author {L.M. Baskin$^c$}
%\author {J. G. Muga$^{a,d}$}
\affiliation{$^a$ Basque Center for Applied Mathematics (BCAM), Alameda de Mazarredo 14, 48009, Bilbao, Spain}
\affiliation{$^b$ Departmento de Qu\'imica-F\'isica, Universidad del Pa\' is Vasco, UPV/EHU, 48009 Leioa, Spain}
\affiliation{$^c$ IKERBASQUE, Basque Foundation for Science, Plaza Euskadi 5, 48009, Bilbao, Spain}
\affiliation{$^{d}$ EHU Quantum Center, Universidad del Pa\' is Vasco, UPV/EHU, 48940 Leioa, Spain}

%\conference{} % An extended version of a conference paper

% Abstract (Do not insert blank lines, i.e. \\) 
\begin{abstract}
	{Uncertainty principle forbids one to determine which of the two paths a quantum system has travelled, unless
		interference between the alternatives had been destroyed by a measuring device, e.g., by a pointer.
		One can try to weaken the coupling between the device and the system, in order to avoid the veto.
		We demonstrate, however, that a weak pointer is at the same time an inaccurate one, and the information about the path 
		taken by the system in each individual trial is inevitably lost. We show also that a similar problem occurs if a classical 
		system is monitored by an inaccurate quantum meter. In both cases one can still determine some characteristic 
		of the corresponding statistical ensemble, a relation between path probabilities in the classical case, and a relation between 
		the probability amplitudes if a quantum system is involved. }
\end{abstract}

\maketitle

\section{Introduction}
There is a well-known difficulty with determining the path taken by a quantum system capable of reaching 
a known final state via several alternative routes. 
According to the Uncertainty Principle~\cite{FeynL}, such a determination is possible only if an additional
measuring device destroys interference between the alternatives. 
However, the~device inevitably perturbs the system's motion, and~alters the likelihood of its arrival at 
the desired final state. The~knowledge of the system's past must, therefore, be incompatible with keeping 
the probability of a successful post-selection intact. 
\par
A suitable measuring device can be a pointer~\cite{vN}, designed to move only if the system travels the chosen path 
so that finding it displaced at the end of experiment could constitute a proof of the system's past. 
{For practical aspects of quantum measurements, see, for~example~\cite{Hybrid_2009,Optomechanics2022}. }
A somewhat naive way around the Uncertainty Principle may be the use of a pointer 
coupled to the system only weakly, thus leaving interference between the paths almost intact.
Perhaps the small change in the pointer's final state {caused by the weak interaction } could provide \e{which path?} (\e{which way?}) 
information previously deemed to be unavailable.
\par
This change can be expressed in terms of a \e{weak value} (WV) \cite{Jordan} of a quantity $\pi_j$, which takes a unit value for the path of interest, 
say, the~path number $j$, and~vanishes otherwise. The~complex valued WV is conveniently defined as the ratio $\la \pi_j\ra_{Weak} \equiv \AA_j/\sum_{i}\AA_i$, 
where $\AA_i$ is the probability amplitude that quantum theory ascribes to the $i$-th path available to the system. {For a quantity $ B $ whose value on the $ i $-th path is $ B_i $, the~WV is $ \expval{B}_W=\sum_i B_i \la \pi_i\ra_W $.} The quantity $\la \pi_j\ra_W$, always known to the theoretician, 
can also be measured by the practitioner. (No surprise here; the response of a quantum system to a small perturbation is usually expressed in terms of probability amplitudes rather than probabilities.) 

{The problem with the just described \e{weak measurements} is to ascribe a physical meaning to a \e{weak value} which is, after~all, a~particular combination of 
	the system's amplitudes. Quantum theory provides only one firm rule: an absolute square $|\AA_j|^2$ yields the relative frequency with which the system travels the $j$-th path, should the interference between the paths be destroyed by a measuring device. But~can there be more rules? For example, what is $\AA_j/\sum_{i}\AA_i$? The mean shift of the \e{weak} pointer. Yes, but~what does it say about the system? The number one obtains dividing $\AA_j$ by the sum of the amplitudes leads to the same final state? Certainly, the~theoretician can figure out this number on the back of an envelope. 
	But what, we insist, does it tell about the system? 
	The question is one of principle: would the knowledge of a WV reveal anything previously unknown about the route
	by which the system reaches its final destination?
}
\par 
The idea is not new, and~was applied, for~example, to~an optical realisation of a three-path problem~\cite{3P1,3P2}. 
The conclusion that the photons can be found in a part of the setup they can neither enter nor leave, and~must therefore have discontinuous trajectories, was subsequently criticised by a number of authors for both technical and more fundamental reasons
{\cite{3Pc1,3Pc3,WV_1,WV_2,WV_3,WV_4,WV_5,WV_6,WV_7,WV_8,WV_9,WV_10,WV_11,WV_12,WV_13,Sokolovski_2023_1,Sokolovski_2023_2}. }
A similar treatment of a four-path \e{quantum Cheshire cat} \cite{4P1,4P3} model suggests the possibility of separating a system from its property, to~wit,
electrons detached from their charges, and~an atom's internal energy \e{disembodied} from the atom itself. 
(For further discussion of the model, the reader is referred to~\cite{4P4}).
The case for a quantum particle (or, at~least, of~some of its \e{properties}) being in several interfering pathways at the same time was recently made in~\cite{Matz}.
\par
Here, our more modest aim is to analyse, in~some detail, the~validity of the approach in the case of the simplest \e{double-slit} (two-path) problem. 
\par 
The rest of the paper is organised as follows. 
Section~\ref{sec:2} briefly describes the well-known quantum double-slit experiment.
A classical analogue of the problem is studied in Sections~\ref{sec:3}--\ref{sec:6}. 
A simple two-way quantum problem is analysed in Sections~\ref{sec:7}--\ref{sec:11}.
Section~\ref{sec:12} contains our~conclusions.

\section{Quantum \e{Which Way?} Problem}\label{sec:2}
\begin{figure}[H]
	\centering\includegraphics[angle=0,width=.4\textwidth]{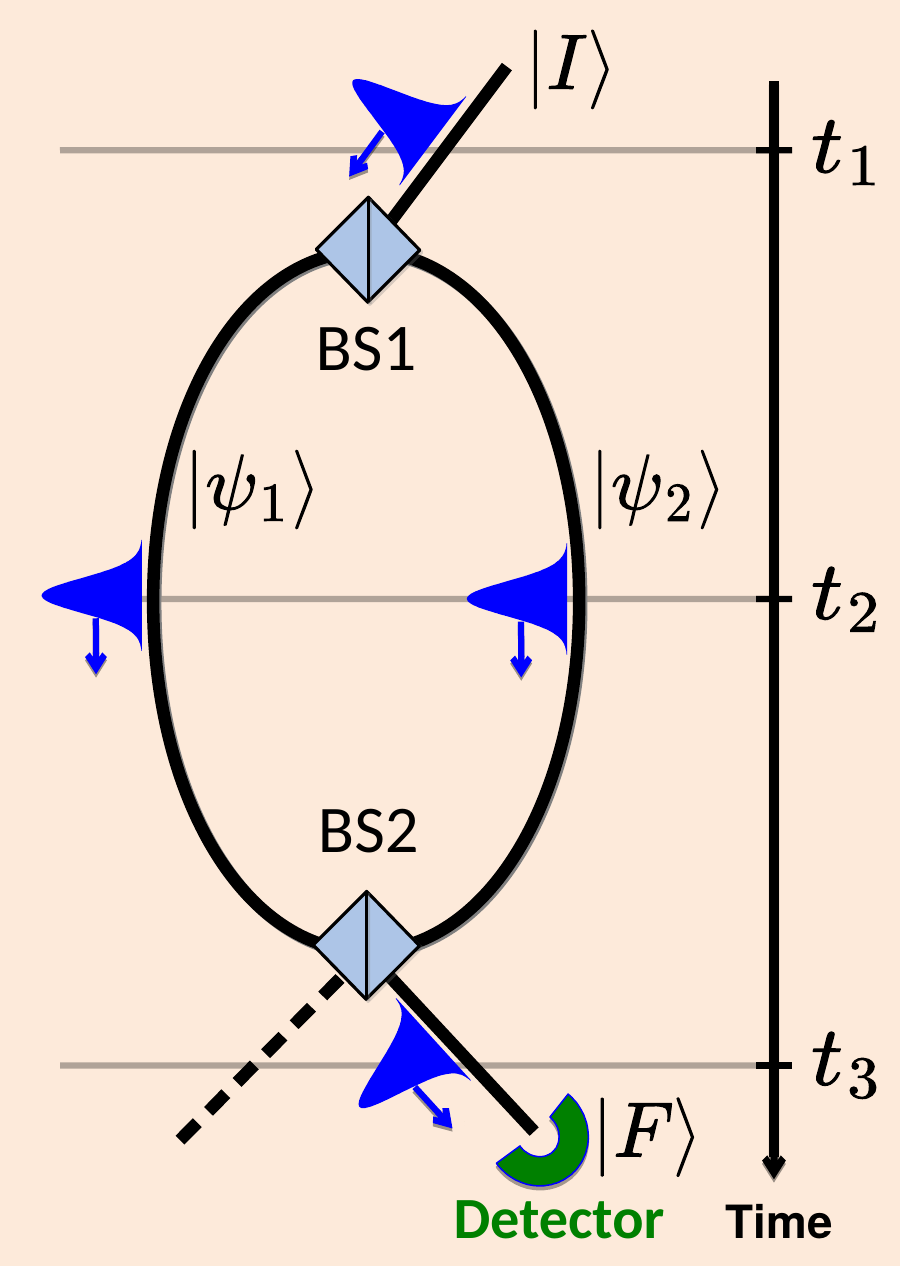}
	\caption{ An optical realisation of a \e{double-slit} experiment.
		At time $ t_1 $, a~photon's wave packet $|I\ra$ is split into two at the first beam splitter (BS1). Its parts 
		$|\psi_1\ra$ and $|\psi_2\ra$ travel
		both optic fibres $ t_2 $ and~are recombined into $|F\ra$ after the second BS2 $ t_3 $. The~two observed events are 
		the initial preparation and~the final detection of the photon. Where was the photon between these two events?
	}\label{F1}
\end{figure}

One of the unanswered questions in quantum theory, indeed its \e{only mystery} \cite{FeynC}, 
concerns the behaviour of a quantum particle in a double-slit experiment shown in Figure~\ref{F1}. 
The orthodox~\cite{FeynL} view is as follows.
With only two observable events, preparation and final detection, 
it is impossible to claim that the particle has gone via one of the slits (paths) and not the other.
This is because the rate of detection by the detector in Figure~\ref{F1} may increase if one of the paths is blocked~\cite{FeynL}.
\par
Neither is it possible to claim that both paths were travelled at the same time 
since an additional inspection never finds only a fraction of a photon in one of the paths~\cite{FeynC}. 

However, such an inspection 
destroys the interference between the paths, and~alters the probability of detection.
The problem is summarised in the Uncertainty Principle~\cite{FeynC}: \e{It is impossible to design any apparatus whatsoever to determine through which hole the particle passes that will not at the same time disturb the particle enough to destroy the~interference}.

A brief digression into Bohmian theory~\cite{Bohm_Hiley} is in order. One can treat the flow lines of a probability density (Bohmian trajectories), obtained from a Schr\"odinger wave function, as~a quantum particle's trajectories. If~so, in~the double-slit case, one finds a single trajectory leading to a given point on the screen and passing through one of the slits. However, a~contradiction with the Uncertainty Principle
stated above is only specious. Bohmian mechanics reproduces all results of the conventional theory, and~the problem of receiving more particles with one slit closed still defies a classical-like explanation. The~comfort of \e{knowing} where the particle was at all times is bought at the price of introducing quantum potential with unusual and potentially non-local properties.
Bohmian trajectories can be evaluated by a theoretician and reconstructed from the \e{weak values} measured by the experimenter
~\cite{Foo2022,Steinberg2016,Strubi_Bruder,Steinberg2011,Wiseman_2007}. 
%The problem is what one should do with these 
The question is how to use or interpret these
trajectories once they have been obtained, in~one way or another. The~consensus appears to be moving away from the original interpretation. Thus, Hiley and Van Reeth suggest~\cite{Hiley_vanReeth} that 
\e{the flow lines $\ldots$ are not the trajectories of single atoms but an average momentum flow}. 
Furthermore, Flack and Hiley~\cite{Hiley_Flack} relate them to Feynman paths (more relevant for our analysis).
Similarly, the~authors of~\cite{Steinberg2016,Steinberg2011} refrain from identifying the average momentum flow lines with individual photon trajectories. Another reason why the Bohmian perspective is of little interest for the present work is because, below, we will limit ourselves to the study of systems in two-dimensional 
Hilbert spaces, where the application of the method is at best problematic. 

\begin{figure}[H]
	\centering\includegraphics[angle=0,width=.45\textwidth]{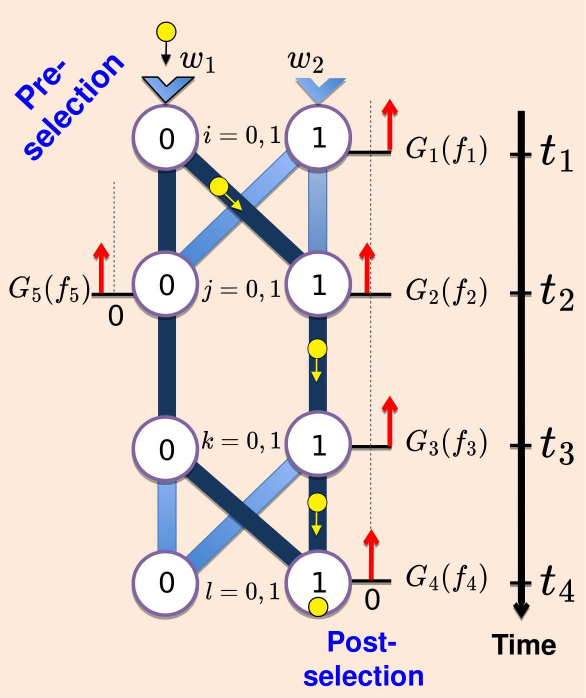}
	\caption{{A classical} %MDPI: We moved figure after the first citation, please confirm. "Perfect, thanks"
		two-state system can reach a final state by taking one of the the eight paths
		with a probability given by Equation~(\ref{a1}). The~system can be monitored with the help of 
		%(possibly inaccurate) 
		pointers (red arrows) which move if the system is detected [cf. Equation~(\ref{a2})].
		%whose initial positions, $f_n$, are distributed according to $G_n(f_n)$ in Equation~(\ref{a2}).
		%A pointer moves by a unit distance whenever the system is observed at the corresponding location.
		Also shown (in dark blue) is the two-way problem of Section~\ref{sec:4}.
	}\label{F2}
\end{figure}

{We return, therefore, to~the possibility of finding a way around the Uncertainty Principle by perturbing the measured system only slightly.
	One such approach, first proposed by Vaidman in Ref.~\cite{3P1}, suggests the following.}
If two von Neumann pointers~\cite{vN}, set up to measure projectors on the paths
(e.g., on~the states $|\psi_1\ra$ and $|\psi_2\ra$ in Figure~\ref{F1}), are coupled to the particle only weakly, interference between the paths can be preserved. 
(One can perform non-perturbing (weak) measurements using von Neumann pointers, either by making coupling to the observed system weak or, equivalently, by~making the pointer's initial position uncertain as we consider.)
If, in~addition, both pointers are found to \e{have moved}, albeit on average, the~\e{weak traces} \cite{3P1} left by the particle will reveal its presence in both paths at the same time. The~idea appears to contradict the Uncertainty Principle and, for~this reason, deserves our attention. We start the investigation by looking first 
at inaccurate pointers designed to monitor a {{classical}} %MDPI: Please confirm if the italics is unnecessary and can be removed. Please check the whole paper and revise if necessary. "Checked and removed"
stochastic system in Figure~\ref{F2}.

%Neither is it possible for the system in Figure~\ref{F2} to make transitions
%$c_1\gets b_0$ or $c_0\gets b_1$, since we ensured $\a(c_k \gets c_j)=\delta_{jk}$
%%%%%%%%%%%%%%%%%%%%%%%%%%%%%%%%%%%%
\section{Consecutive Measurements of a Classical~System}\label{sec:3}
Our simple classical model is as follows. (We ask for the reader's patience. The~quantum case will be discussed shortly.)
% (We ask for the reader's patience. Our aim is to to distinguish between the features 
%common to a classical and quantum analyses, and those which are purely quantum.)
A system (one can think of a little ball rolling down 
a network of connected tubes shown in Figure~\ref{F2}) is introduced into one of the two inputs at $t=t_1$, with~a probability $w_i$, $i=0,1$. It then passes through states {$j$ and $k$, where $j,k=0,1$ }
at the times $t_2$, and~$t_3$, respectively. The~experiment is finished when the system is collected 
in a state $l$, $l=0,1$ at $t=t_4$. From~each state $i$, the~system is directed to one of the states $j$ with a probability $p(j\gets i)$,
similarly from $j$ to $k$, and~finally from $k$ to $l$. There are altogether eight paths $\{l\gets k\gets j \gets i\}$, each travelled with a {probability} %MDPI: 1. Please make sure the variables format in equations and text are unified (italic or non-italic, bold or not, subscript/superscript or not). Please check and revise if necessary. 2. Please check all the equations and make sure no repeated equations exist. "Checked and confirmed"
\begin{align}\label{a1}
&P(l\gets k\gets j \gets i)= p(l\gets k)p(k\gets j)p(j\gets i),\\
&p(k\gets j)=\delta_{jk},\nonumber
\end{align}
where $\delta_{jk}$ is the Kronecker delta.
(The choice of this design will become clear shortly.)

We make the following~assumptions.

\begin{enumerate}
	\item {Alice, the~experimenter, knows the path probabilities in Equation~(\ref{a1}) but~not the input values $w_i$.}
	\item {She cannot observe the system directly, and~relies on the readings of pointers with positions $f_n$, $n=1,2,\ldots,5$, installed at different locations as~shown in Figure~\ref{F2}.
		If the system passes through a location, the~corresponding pointer is displaced by a unit length, $f_n\to f_n+1$; otherwise, it is left intact. }
	\item {The pointers are, in~general, inaccurate since their initial positions are distributed around zero with probabilities 
		$G_n(f_n)$ (see Figure~\ref{F2}). Their final positions are, however, determined precisely.} 
	We will consider the distributions $G_n$ to be Gaussians of widths~$\dd_n$:
	\begin{align}\label{a2}
	&G_n(f_n)=\sqrt{\frac{1}{\pi(\dd_n)^2}}\exp\left (-\frac{f_n^2}{(\dd_n)^2}\right ),\\
	&\int G_n(f_n) df_n=1, \q G_n(f_n){\xrightarrow[ \dd_n \to 0 ] {}} \delta(f_n).\nonumber
	\end{align}
\end{enumerate}
%\vspace{1mm}
{The experiment}  ends just after $t=t_4$, when Alice's observed outcomes are the
five numbers $f_n$, $n=1,\dots,5$.
These are distributed with a probability density\vspace{-10pt}

\begin{widetext}
\begin{align}\label{a3}
	\rho(f_1,f_2,f_3,f_4,f_5)= \sum_{i,j,k,l=0,1}w_i P(l\gets k\gets j \gets i)\times
	G_1(f_1-i)G_2(f_2-j)G_3(f_3-k)G_4(f_4-l)G_5(f_5+j-1).
\end{align}
\end{widetext}

%Alice cannot verify the prediction 
Equation (\ref{a3}) is not particularly useful 
%Bob the theorist to estimate the probabilities $\rho(f_1,f_2,f_3,f_4,f_5)$.}
since $w_i$ %in Equation~(\ref{a3}) 
are unknown. However, by~making 
the first pointer accurate, $\dd_1 \to 0$, $G_1(f_1)\to \delta (f_1-i)$ where $\delta(x)$ is the Dirac delta, she is able to
{\it pre-select} those cases, where, say, $f_1= 0$, and~collect only the corresponding statistics. 
%Now if $G_1(f_1)\ne 0$, $G_1(f_1-1)\equiv 0$, and the new normalised distribution 
%$\rho_0(f_2,f_3,f_4,f_5) \equiv \rho(f_1= 0,f_2,f_3,f_4,f_5)/\int\rho(f_1=0,f_2,f_3,f_4,f_5)$ $df_2df_3df_4df_5$
Now the (properly normalised) distribution of the remaining four readings 
does not depend on $w_i$,%\vspace{-10pt}
\begin{align}\label{a4}
\rho_0(f_2,f_3,f_4,f_5)= \sum_{j,k,l=0,1}P(l\gets k\gets j \gets 0)\times&\\
G_2(f_2-j)G_3(f_3-k)G_4(f_4-l)G_5(f_5+j-1)&,\nonumber
\end{align}
%\noindent
and Alice has a complete description of the pre-selected ensemble.
\par
{Alice can also {\it post-select} the system by selecting, for example,~the cases where it ends in a state $1$ at $t=t_4$.
	With $G_4(f_4) \to \delta(f_4-1)$, the~remaining random variables $f_2$, $f_3$, and~$f_5$ are distributed according to
	[cf. Figure~\ref{F2} and Equation~(\ref{a1})]
	\begin{align}\label{a5}
	&\rho_{1\gets 0}(f_2,f_3,f_5)=\\
	% \q\q\q\n
	%\q\q\q\q\q\q\q\q\q\q\q\q\n 
	% \bigg[ 
	&\frac
	{ \sum_{j=0,1}P_j G_2(f_2-j)G_3(f_3-j)G_5(f_5+j-1)}
	%\bigg ]\n
	%\times \bigg[
	{P_0+P_1},\nonumber
	%\bigg ]^{-1}\q\q\
	\end{align}}
where we introduce a shorthand
\begin{align}\label{a6}
P_j\equiv P(1\gets j \gets j\gets 0), \q j=0,1.
\end{align}
{Equation}~(\ref{a4}) suggests a simple, yet useful, general~criterion. 
\begin{itemize}
	\item { Alice can determine the system's past location {\it only} 
		when she obtains a pointer's reading
		% (or a set of readings) 
		whose likelihood depends {\it only} on the 
		%system's
		probabilities of the system's paths passing through that location.} 
	%In Equation~(\ref{a5}) such are the four paths passing through $i=0$ and $l=1$. }
\end{itemize}

{{For example}, at~least three accurate readings
	($f_1$, $f_4$ and one of $f_2$,$f_3$ or $f_5$,) 
	are needed if Alice is to know which of the eight paths shown in Figure~\ref{F2} the system has travelled
	during each trial. With~$G_n(f_n)=\delta(f_n)$ for $n=1,2,4$, a~trial can yield, for example,~
	the values $f_1=0$, $f_2=1$, and~$f_4=1$. The~likelihood of these outcomes is given by the probability $P(1\gets 1\gets 1 \gets 0)$ in Equation~(\ref{a1}),
	and Alice can be certain that the route $\{1\gets 1\gets 1\gets 0\}$ has indeed been travelled. }
%%%%%%%%%%%%%%%%%%%%%%%%
\section{A Classical \e{Two-Way~Problem}}\label{sec:4}
Consider next a pre- and post-selected ensemble 
%described by Equation~(\ref{a3}),
with two routes connecting the states $0$ at $t=t_1$ and $1$ at $t=t_4$ (shown in Figure~\ref{F2} in dark blue).
%\begin{eqnarray}\label{h1}
%P(1\gets 1 \gets 1\gets 0)=P(1\gets 0 \gets 0\gets 0).
%\end{eqnarray}
As a function of the second pointer's accuracy, $\dd_2$, the~distribution of its readings (\ref{a5}) changes from a bimodal, when the pointer is accurate
\begin{align}\label{h1a}
\rho_{1\gets 0}(f_2)&\equiv \int \rho_{1\gets 0}(f_2,f_3,f_5)df_3 df_5\nonumber\\
&=\frac{P_0G_2(f_2)+P_1G_2(f_2-1)}{P_0+P_1}\nonumber\\
&{\xrightarrow[ \dd_2 \to 0] {}}\frac{P_0\delta(f_2)+P_1\delta(f_2-1)}{P_0+P_1}
\end{align}
to the original broad Gaussian for an inaccurate pointer,
\begin{align}\label{h1b}
\rho_{1\gets 0}(f_2)
{\xrightarrow[ \dd_2 \to \infty ] {}} \thinspace\thinspace G_2(f_2-z),\q\q\q
\end{align}
displaced as a whole by
\begin{align}\label{h1c}
z=\frac{P_1}{P_0+P_1}.
\end{align}
%{\rrr For a visual representation of this transition, see Figure~\ref{F6} a).}

Equation~(\ref{h1b}) reflects a known property of Gaussians, to~our knowledge, first explored in~\cite{Vaid}, and~discussed in detail in Appendix \ref{appx:A}. The~transformation of two peaks (\ref{h1a}) into a single maximum (\ref{h1b}) is best described by the catastrophe theory~\cite{Cata}. For~example, for~$P_0=P_1$, 
a pitchfork bifurcation converts two maxima and a minimum into a single maximum for $\dd_2=\sqrt{2}$ {(see Figure~\ref{F6}(\textbf{a}) of Appendix \ref{appx:B})}. 
\par
With a sufficiently accurate pointer $\dd_2\ll1$, a~reading always lies close to $0$ or $1$, and~in every trial, Alice knows the path followed by the system.
%\cite{FOOT}. 
%This agrees with the rule of the previous Section, since the probability of obtaining $f_2=1$ and $f_4=1$, $\rho_0(f_2=1,f_4=1)=\int \rho_0(1,f_3,1,f_5)df_3 df_5=P(1\gets 1\gets 1 \gets 0)$, depends only on the probability of travelling the path $\{ 1\gets 1\gets 1 \gets 0\}$
\par
With a highly inaccurate pointer $\dd_2\gg 1$, not a single reading $f_2$ 
can be attributed to one path in preference to the other, and~the route by which the system arrived at its final state is never known (see Appendix \ref{appx:C}). 
%{\r As any pointer reading contains contributions from both alternatives.}
Indeed, for~$P_0=P_1$, even the most probable outcome $f_2=1/2$ is equally likely to occur if the system takes path $\{ 1\gets 0\gets 0 \gets 0\}$, 
or $\{ 1\gets 1\gets 1 \gets 0\}$,
\begin{align}\label{h1d}
\rho\left(f_2=\frac{1}{2}\right)={ \frac{1}{2}}
\left[G_2\left(\frac{1}{2}\right)+G_2\left(-\frac{1}{2}\right)\right], 
\end{align}
and the \e{which way?} information is clearly lost.
%As $\dd_2\to \infty$, Alice looses all information about the path travelled by the system in any given trial. 
\par
Still, something can be learned about a pre- and post-selected classical ensemble, even without knowing the path taken by the system. 
Having performed many trials, Alice can evaluate an average reading,
\begin{align}\label{h2}
\la f_2\ra\equiv \int f_2\rho_{1\gets 0}(f_2) df_2 = z.
%\frac{P(1\gets j \gets j\gets 0)}{\sum_{j=0,1}P(1\gets j \gets j\gets 0)}=1/2
\end{align}
{The quantity} $z$ in Equations~(\ref{h1c}) and (\ref{h2}) is the relative (i.e., renormalised to a unit sum) probability
of travelling the path $\{ 1\gets 1\gets 1 \gets 0\}$, and~is independent of $\dd_2$. 
Thus, by~using an inaccurate pointer, Alice can still estimate certain parameters of her statistical~ensemble. 
%Alice can always measure the path probabilities (it will take more trials of $\dd_2$ is large), 
%even if she is unable to answer the \e{which way?} question. 
%{\r
%Given the frequentist interpretation of probabilities, Alice (the experimenter) even without having prepared the experiment herself, say Bob (the theoretician) did so, she could learn the probabilitites of each path after repeating many trials.}
%%%%%%%%%%%%%%%%%%%%%%%%%%%%%%%%%%%
\section{Two Inaccurate Classical Pointers and a Wrong~Conclusion}\label{sec:5}
\begin{figure}[H]
	\centering\includegraphics[angle=0,width=0.4\textwidth]{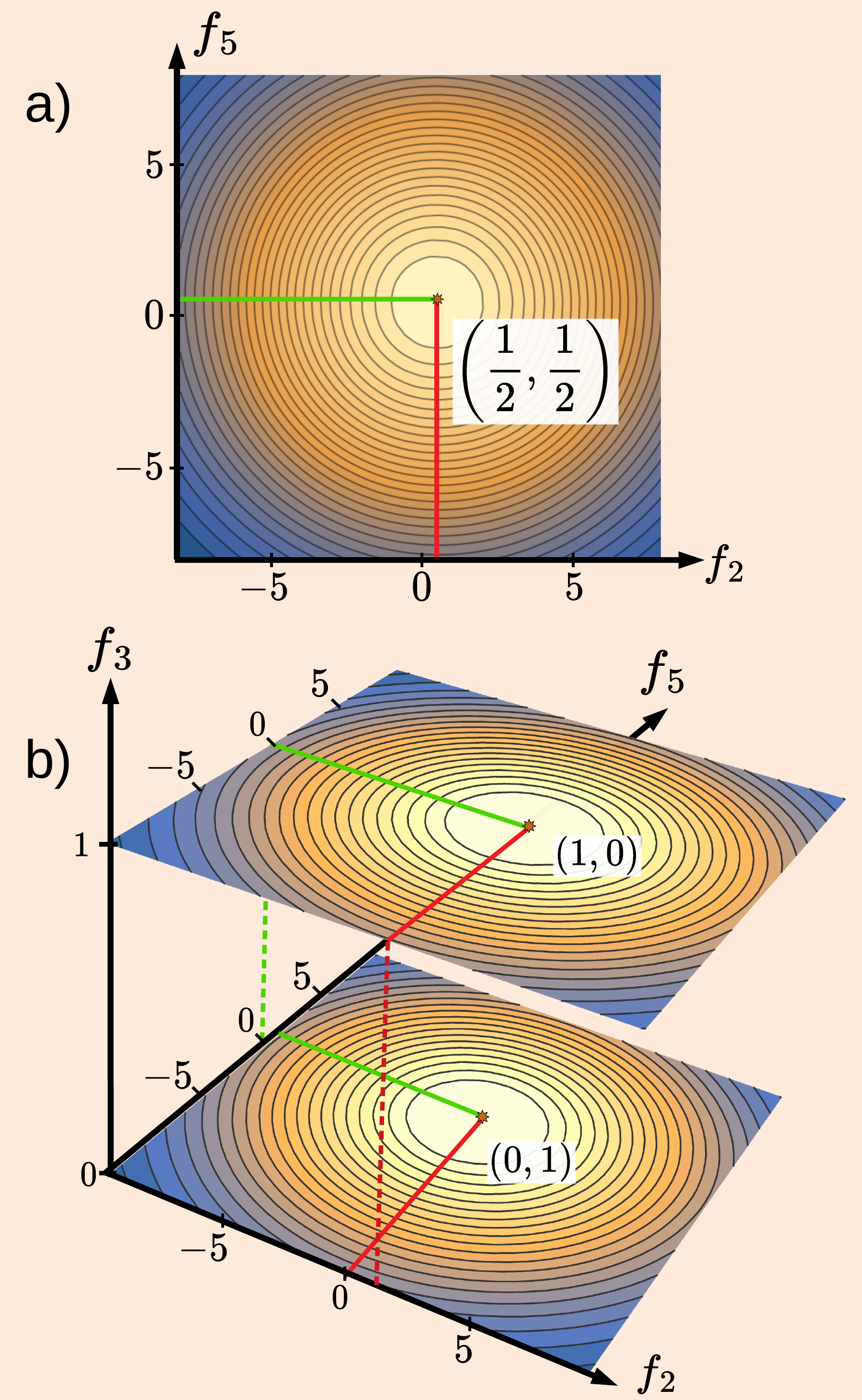}
	\caption{(\textbf{a}) {Distribution of} %MDPI: Please add the left bracket in the image, e.g., “a)” should be “(a)”. "Checked"
		the readings $\rho_{1\gets 0}(f_2,f_5)$ of two inaccurate pointers with $P_0=P_1$ and $\dd_2=\dd_5=10$, monitoring a classical system at $t=t_2$ [cf. Equation~(\ref{f1})]; (\textbf{b}) the distribution $\rho_{1\gets 0}(f_2,f_3,f_5)$ in Equation~(\ref{h3})
		with an accurate pointer, $\dd_3\to 0$, added at $t=t_3$. Note that the integration of the distribution in ({b}) over $df_3$ 
		recovers the distribution shown in ({a}). }
	\label{F3}
\end{figure}

A word of caution should be added against an attempt to recover the \e{which way?} information 
with the help of Equation~(\ref{h2}).
% It could go like this. 
For two equally inaccurate pointers,\linebreak $\dd_2 =\dd_5 \equiv \dd \gg 1$ [cf. Figure~\ref{F2}],
the distribution of the readings tends to a single Gaussian shown in Figure~\ref{F3}(\textbf{a}) (see also Appendix \ref{appx:A}), \vspace{-10pt}
%peaked at $(f_2=1/2, f_5=1/2)$,

	%\centering %% If there is a figure in wide page, please release command \centering
\begin{eqnarray}\label{f1}
\rho_{1\gets 0}(f_2,f_5)=\q\q\q\q\q\q\q\q\q\q\q\q\n
\frac{P_0G_2(f_2)G_5(f_5-1)+P_1G_2(f_2-1)G_5(f_5)}{P_0+P_1}\q\q\q\n
{\xrightarrow[ \d\to \infty ] {}}\thinspace G_2(f_2-z_2)G_5(f_5-z_5),\q\q\q\q\q
% (\pi \d^2)^{-1}\exp\left [-\frac{(f_2-1/2)^2}{\d^2}-\frac{(f_5-1/2)^2}{\d^2}\right].
\end{eqnarray}

where
\begin{align}\label{f2}
z_2=\frac{P_1}{P_0+P_1}, \q z_5=\frac{P_0}{P_0+P_1}=1-z_2.
\end{align}
%\newpage

It may seem that (the reader can already see where we are going with this) the following hold:
\begin{enumerate}
	\item[(i)]{Each pointer in Equation~(\ref{f1}) \e{moves} only when the system is in its path.}
	\item[(ii)]{Equation~(\ref{f1}) suggests that both pointers have moved (albeit on average). }
	\item[(iii)]{Hence, the~system must be travelling both paths at the same time.}
\end{enumerate}
%The last conclusion is clearly wrong, since the system in Figure~\ref{F1} does not split in two. 
%It is also wrong, because the \e{which way?} information, lost in each individual reading, cannot be recovered 
%from the distribution's peak, or from its first moment, $z_2$, or $z_5$. 

{To check} if this is the case, Alice can add an accurate pointer ($f_3$) acting at $t=t_3$ (see Figure~\ref{F2}).
If parts of the system were in both places at $t_2$, the~same must be true at $t_3$ since Alice
% {\r [or Bob, the theoretician that designed the experiment]} 
makes sure that 
no pathway connects the points $j=0$ and $k=1$. The~accurate pointer should, therefore, always find only a part of the 
system. Needless to say, this is not what happens. With~an additional accurate pointer in place, three-dimensional distribution (\ref{a5}) becomes bimodal, {again} (see Figure~\ref{F3}(\textbf{b}))%\vspace{-10pt}
%in the three dimensional space,
\begin{align}\label{h3}
\rho_{1\gets 0}(f_2,f_3,f_5)=(P_0+P_1)^{-1}\times&\\
%\q\q\q\q\q\q\q\q\q\q\q\q\q\q\n
\big[P_0G_2(f_2)G_5(f_5-1)\delta(f_3)+&\nonumber\\
P_1G_2(f_2-1)G_5(f_5)\delta(f_3-1)\big]&.\nonumber
%{\xrightarrow[ \d \to \infty ] {}} \n
%  (\pi \d^2)^{-1}\exp\left [-\frac{(f_2-1/2)^2}{\d^2}-\frac{(f_5)^2}{\d^2}\right].
\end{align}
{An inspection} of statistics collected separately for $f_3=0$ or $f_3=1$ shows that at $t=t_2$, only one of the two pointers 
{moves} in any given trial. 
Contour plots of the densities in Equations~(\ref{f1}) and (\ref{h3}) are shown in Figure~\ref{F3}a,b, respectively.
The fallacy (i)--(iii), evident in our classical example, will become less obvious in the quantum case we will study after a brief~digression.

%when given an extra dimension, i.e.,~projecting in a given way, inconsistencies disappear.

%\begin{figure}[ht]
% \includegraphics[angle=0,width=.2\textwidth]{Penrose.jpg}
% \caption{ {Hossenfelder S and Palmer T (2020) Rethinking Superdeterminism. Front. Phys. 8:139. doi: 10.3389/fphy.2020.00139} }\label{fig:Penrose}
%\end{figure}

%%%%%%%%%%%%%%%%%%%%%%%%%%%%%%%%/Users/dmitri/Desktop/ANTON_WV/Fig.3d_class 2.pdf
\section{Classical \e{Hidden~Variables}}\label{sec:6}
Before considering the quantum case, it may be instructive to add a fourth assumption to the list of Section~\ref{sec:3}.
\begin{enumerate}
	\setcounter{enumi}{3}
	\item In Alice's world, all pointers have the property that an accurate detection inevitably perturbs the system's evolution. 
\end{enumerate}

{For example}, whenever the pointer $f_3$ moves, the~probabilities $p(l\gets1)$ in Equation~(\ref{a1}) are reset to
$p'(l\gets1,\dd_3)$. Thus, $P_1$ changes to $P'_{1}(\dd_3$),
%$P_{1\gets 0}(\dd_3$), 
while $P_0$ remains the same. 
The change may be the greater the smaller $\dd_3$ is and~$P'_{1}(\dd_3\to \infty)=P_1$. 
Now the system, accurately observed in the state $1$ at $t_3$ arrives 
in state $1$ at $t=t_4$, say, less frequently than it would with no pointer ($f_3$) in place, 
%$P_{1\gets 0}(\dd_3 \to 0)= 
$P_0+P'_1(\dd_3 \to 0) < 
P_0+P_1$.
%\par
So, where was the {\it unobserved} system at $t=t_3$?
%arriving at $t_4$ in $1$ with a probability $P_{1\gets 0}(\dd_3\to \infty)= P_0+P_1(\dd_3 \to \infty)$ 
\par
Empirically, the~question has no answer. To~ensure the arrival rate is unchanged by observation, Alice can only use an inaccurate pointer, $\dd_3\ \gg 1$,
which yields no \e{which way?} information. Performing many trials, she can, however, measure both 
the probability of arriving in $1$ at $t=t_4$, $W_1(t_4)=P_0+P'_1(\dd_3 \gg 1)\approx P_0+P_1$ and the value of
$z=P'_1(\dd_3 \gg 1)/[P_0+P'_1(\dd_3 \gg 1)]\approx P_1/[P_0+P_1]$ [cf. Equations~(\ref{h1b}) and (\ref{h1c})].
She can then evaluate unperturbed path probabilities,
\begin{align}\label{cc0}
P_1\approx zW_1(t_4), \q P_0\approx (1-z)W_1(t_4).
\end{align}
{Having observed} that $P_0$ and $P_1$ are both positive, and~do not exceed unity, Alice 
may reason about what happens to the unobserved system in the following manner. 
The available empirical data are 
consistent with the system, always following one of the two paths with probabilities in Equation~(\ref{cc0}). 
However, with~the available instruments, it is not possible to verify this conclusion experimentally. 
\par
This is as close as we can get to the quantum case using a classical toy model. 
We consider the quantum case~next.

\section{Consecutive Measurements of a~Qubit}\label{sec:7}
\begin{figure}[ht]
	\centering\includegraphics[angle=0,width=0.48\textwidth]{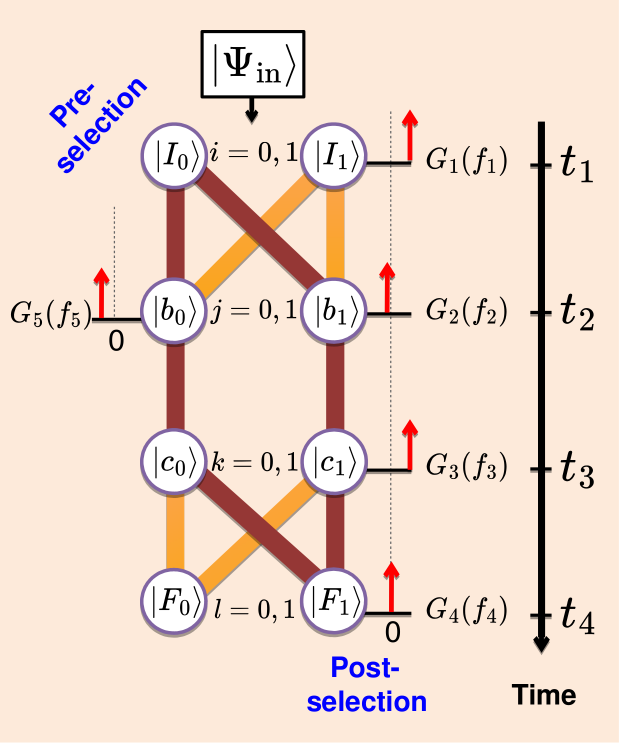}
	\caption{ A two-level quantum system can reach final states $|F_l\ra$, $l=0,1$, via eight virtual paths
		whose probability amplitudes are given by Equation~(\ref{c0}). The~system is monitored by means of
		%with the help of (possibly inaccurate) 
		von Neumann pointers (red arrows), set to measure projectors $\pin_n$ [cf. Equation~(\ref{c1})].
		Also shown (in maroon) is the \e{double-slit problem} of Section~\ref{sec:8}.
		% for a system pre- and post-selected in the states $|I_0\ra$ and $F_1\ra$.
		%is marked by black borders. 
	}\label{F4}
\end{figure}

A quantum analogue of the classical model just discussed is shown in Figure~\ref{F4}.
An experiment in which Alice monitors the evolution of a two-level quantum system (qubit) 
with a Hamiltonian $\h^{s}$ by means of five von Neumann pointers
begins at $t=t_1$ and ends at $t=t_4$. 
With no transitions between the states $|b_{0(1)}\ra$ and $|c_{1(0)}\ra$, 
there are altogether eight virtual (Feynman) paths which connect the initial and final states. 
Just before $t_1$, the qubit may be thought to be in some state $|\Psi_{\text{in}}\ra$, and~the eight path amplitudes 
are given by ($i,j,k,l=0,1$) [cf. Figure~\ref{F4}]
\begin{align}\label{c0}
&\A(F_l \gets c_j \gets b_j\gets I_i) =\\
&\a(F_l\gets c_j)\a(c_j\gets b_j)\a(b_j\gets I_i),
%\times \la I_i|\Psi_{in}\ra,\q\q\q\q\q\q\q\q\q\q\q\q
\end{align}
where $\a(b_j \gets I_i)=\la b_j|\hat U^s(t_2-t_1)|I_i\ra$, etc., and~$\hat U^s(t)\equiv\exp(-i\hat H^st)$ is the qubit's own evolution~operator. 

\vspace{.3cm}
We note the~following.
\begin{adjustwidth}{-.5cm}{-1cm}
\begin{enumerate}
	\item {Alice the experimenter knows the path amplitudes in Equation~(\ref{c0}) but~not the system's input state $|\Psi_{\text{in}}\ra$.
		(If she did, the~experiment would begin earlier, at~the time $|\Psi_{\text{in}}\ra$ was first determined.) }
	\item {Alice cannot look at the system directly and~has access only to von Neumann pointers~\cite{vN},
		with positions $f_n$, and~momenta $\lm_n$, $n=1,\ldots, 5$ (see Figure~\ref{F4}).
		The pointers
		%, whose positions and momenta are $f_n$ and $\lm_n$, respectively, 
		are briefly coupled
		to the system at $t=t_n$, ($t_5\equiv t_2$), via
		\begin{align}\label{c1}
		\h^{\text{int}}_n =-i\partial_{f_n}\pin_n \delta(t-t_n), 
		\end{align}
		where
		\begin{align}
		\begin{split}
		\pin_1&=\dyad{I_1},\th
		\pin_2=\dyad{b_1},\th
		\pin_3=\dyad{c_1},\\
		\pin_4&=\dyad{F_1},\th
		\pin_5=\dyad{b_0}
		\end{split}
		\end{align}
		%\hat \lm_n \pin_n \delta(t-t_n), \q \pin_n=|X_n\ra\la X_n|,\q \n
		and have no own dynamics. }
	\item {The pointers, initially in states $|G_n\ra$, are inaccurate, with~initial positions distributed around zero with probability amplitudes 
		$G_n(f_n)\equiv \la f_n|G_n\ra$.
		%although their final positions are determined precisely. 
		We consider Gaussian pointers,
		% $|G_n\ra$, $n=1,...5$,
		\begin{align}\label{c2}
		&G_n(f_n) =\left(\frac{2}{\pi(\dd_n)^2}\right)^{1/4} \exp\left (-\frac{f_n^2}{(\dd_n)^2}\right),\\
		&\int G^2_n(f_n) df_n=1,
		\q\q G^2_n(f_n){\xrightarrow[ \dd_n \to 0 ] {}} \delta(f_n).\nonumber
		%\q\q\q\q\q\q\q\q\q\q
		\end{align}}
	%where $G_n(f_n)$ are now probability amplitudes rather than probabilities. }
	\item {A pointer perturbs the qubit's evolution, except~in the limit $\dd_n\to \infty$.
		Indeed, replacing $f_n$ with $f_n'=f_n/\dd_n$ changes $\h^{\text{int}}$ in Equation~(\ref{c1}) to $\h^{\text{int}'}=\h^{\text{int}}/ \dd_n$, 
		and a highly inaccurate pointer decouples from the qubit~\cite{4P4}. Vice~versa, 
		a weakly coupled pointer is, necessarily, an~inaccurate one.
	}
\end{enumerate}
\end{adjustwidth}

As in the classical case, to~be able to make statistical predictions, Alice needs to make the first measurement accurate, $\dd_1\to 0$, 
$G^2_1(f_1)\to \delta(f_1)$, and~pre-select, for example,~only those cases where $f_1=0$, 
thereby preparing the system 
%and the system is thereby prepared 
in the state $|I_0\ra$. 
%\yy{Should include something about preparation?} 
The rest of the readings are distributed according to (we use $\tilde \rho$ to distinguish
from the classical distributions of Sections~\ref{sec:3}--\ref{sec:5})
\begin{widetext}
\begin{align}\label{c4}
\tilde\rho_0 (f_2,f_3,f_4,f_5)=\sum_{l=0,1}|G_4(f_4-l)|^2\times
%\q\q\q\q\q\q\q\q\q\q\q\q\n
%\equiv \la \Phi_F|f_a\ra|f_b\ra|f_c\ra|f_d\ra \la f_d|\la f_c|\la f_b|\la f_a|\Phi_F\ra=\q\q\q\q\q\q\n
\bigg |\sum_{j=0,1} G_3(f_3-j)G_2(f_2-j) 
G_5(f_5+j-1) \A{(F_l\gets c_j \gets b_j\gets I_0)}\bigg|^2.
%+\bigg|G_4(f_4)\sum_{i,j,k=0,1} G_3(f_3-k)G_2(f_2-j)G_1(f_1-i) \n
%\times G_5(f_5+j-1) \A^s{(F_0\gets c_k \gets b_j\gets I_i)}\la I_i|\Psi_{in}\ra\bigg|^2\
\end{align}
\end{widetext}
{As in} the classical case, Alice can also post-select the qubit, e.g.,~in a state $|F_1\ra$, 
by choosing $ \dd_4 \to 0$, $G^2_4(f_4)\to \delta(f_4)$, and~collecting the statistics only if $f_1=0$ and $f_4=1$.
The distribution of the remaining three readings is given by
\begin{eqnarray}\label{c5}
\tilde\rho_{1\gets 0} (f_2,f_3,f_5)= 
%\mathcal{N}
%P_{1\gets 0}
W_1(t_4)
%(\d_2,\d_3\d_5)
^{-1}\times\q\q\q\n
\bigg |\sum_{j=0,1} G_3(f_3-j)G_2(f_2-j)
% \q\q \n 
G_5(f_5+j-1) \A_j
% \A{(F_1\gets c_j \gets b_j\gets I_0)}
\bigg|^2,
\end{eqnarray}
where we introduce a shorthand
\begin{align}\label{c6}
\A_j\equiv \A(F_1\gets c_j \gets b_j\gets I_0), \q j=0,1.
\end{align}
{The normalisation} factor $W_1(t_4)\equiv \int \rho_{1\gets 0} (f_2,f_3,f_5)$ $df_2 df_3 df_5$ is the probability of reaching the final state $|F_1\ra$
with all three pointers in place, 
which depends on the pointers' accuracies 
%since the qubit is perturbed by the measurements,
%($n=2,3,5$)
%+\bigg|G_4(f_4)\sum_{i,j,k=0,1} G_3(f_3-k)G_2(f_2-j)G_1(f_1-i) \n
%\times G_5(f_5+j-1) \A^s{(F_0\gets c_k \gets b_j\gets I_i)}\la I_i|\Psi_{in}\ra\bigg|^2\
%\end{eqnarray}
\begin{align}\label{c6}
% P_{1\gets 0}
W_1(t_4)
&=|\A_0|^2+|\A_1|^2+ 2J_2 J_3 J_5 
\R\left [{\A_0}^* \A_1\right ],\\% \q \q\n
J_n &\equiv \int G_n(f_n)G_n(f_n-1)df_n=
\exp\left(\frac{-1}{2(\dd_n)^2}\right).\nonumber
\end{align}
{The general} rule of the previous section can be extended to the quantum case as~follows. 
\begin{itemize}
	\item { 
		Alice may ascertain the qubit's condition, represented by a state in its Hilbert space, {\it only} 
		when she obtains a pointer's reading 
		%(or a set of readings) 
		whose probability depends {\it only} on the system's path amplitudes for the paths passing through the state in question. }
	%consistent with the condition. {\r See Appendix \ref{label}.}}
	%In Equation~(\ref{c5}) such are the two paths connecting $|I_0\ra$ with $|F_1\ra$. }
\end{itemize}

{As in} the classical case, three accurate measurements allow one to determine the path followed by the qubit. 
For example, with~$\dd_1,\dd_2, \dd_4\to 0$, outcomes $f_1=0$, $f_2=1$, and $f_4=1$, whose probability is
\begin{align}\label{c7}
%\tilde\rho_0 (f_2,f_4
P(1,1,0) \equiv \int_{-\ep}^{\ep} df_2  \int_{1-\ep}^{1+\ep} df_4\int_{-\infty}^{\infty}df_3 df_5 &\\
\times \tilde\rho_0 (f_2,f_3,f_4,f_5) 
{\xrightarrow[ \d_2,\thinspace\d_4 \to 0 ] {}}|\A_1|^2&,\nonumber
\end{align}
indicates that the qubit has followed the path $\{F_1\gets c_1 \gets b_1\gets I_0\}$ (see Figure~\ref{F4}).
%A similar result can be obtained also by sending $\dd_3$ or $\dd_2$ to zero. 
%With the pre- and post- selections made, it takes one more accurate reading to distinguish between all four of the qubit's paths.
%In each trial the path is known, 
%and Alice recovers
%the classical setup shown in Figure~1, with the path { probabilities} given by $P(\l\gets j\gets j \gets 0)= |\A(\l\gets j\gets j \gets 0)|^2$. 
%%%%%%%%%%%%%%%%%%%%%%%%%%%
\section{A Quantum \e{Double-Slit} Problem}\label{sec:8}
The simple model shown in Figure~\ref{F4} has the essential features of the setup shown in Figure~\ref{F1} and~is simple to analyse.
Two paths connect the initial and final states, $|I_0\ra$ at $t=t_1$ and $|F_1\ra$ at $t=t_4$; pointers $f_2$ and $f_5$
monitor the presence of the qubit in each path at $t=t_2$, and~the pointer $f_3$ can be used for additional control.
%Suppose next that Alice studies a pre-and post-selected ensemble described by Equation~(\ref{c5}), with two routes connecting the states $|I\ra$ at $t=t_1$ and $|F_1\ra$ at $t=t_4$,
For simplicity, Alice can decouple two pointers from the qubit by sending
\begin{align}\label{g1}
\dd_3,\dd_5\to \infty, \q J_3,J_5\to 1.
\end{align}
{As a} function of the remaining pointer's accuracy $\dd_2$, the~distribution of its readings (\ref{c5}) changes from bimodal,
\begin{eqnarray}\label{g2}
\rr_{1\gets 0}(f_2)\equiv \int \rr_{1\gets 0}(f_2,f_3,f_5)df_3 df_5\q\q\n
{\xrightarrow[ \d_2 \to 0] {}} \frac{|\A_0|^2\delta(f_2)+|\A_1|^2\delta(f_2-1)}{|\A_0|^2+|\A_1|^2},
% {\xrightarrow[ \d \to \infty ] {}} \q G_2(f_2-1/2).\q\q\q
\end{eqnarray}
to a single broad Gaussian, 
\begin{eqnarray}\label{g4}
\rr_{1\gets 0}(f_2)=
\frac{|G_2(f_2)\A_0+G_2(f_2-1)\A_1|^2}{|\A_0|^2+|\A_1|^2+ 2  
	\R\left [{\A}^*_0 \A_1\right ]}\n
{\xrightarrow[ \d_2 \to \infty ] {}} \q G^2_2(f_2-\zz),\q\q\q
\end{eqnarray}
displaced as a whole by
\begin{align}\label{g3}
\zz=\R\left [\frac{\A_1}{\A_0+\A_1}\right], 
\end{align}
where we use Equation~(\ref{xx4}) of Appendix \ref{appx:D}.
The transformation {between the two forms is similar to transformation of the classical probability from (\ref{h1a}) to (\ref{h1b}) (see Appendix \ref{appx:D}).}
%Equation (\ref{h1b}) reflects a curious property of Gaussians, to our knowledge first discovered in~\cite{Vaid}, and discussed in details in Appendix A. . 
\par
As in the classical case, with~an accurate pointer $\dd_2\ll1$, a~reading is always either $0$ or $1$, and~in every trial, Alice knows the path followed by the qubit. 
%This agrees with the criterion of the previous Section, since the probability of obtaining $f_2=0$ and $f_4=0$, $\rho_0(f_2=1,f_4=1)=\int \rho_0(1,f_3,1,f_5)df_3 df_5=|\A_1|^2$, depends only on the probability amplitude of the path $\{ 1\gets 1\gets 1 \gets 0\}$.

{For a highly} inaccurate pointer $\dd_2\gg 1$, there is not a single reading $f_2$ which can be attributed to one path in preference to the other (cf. Appendix \ref{appx:C}), so Alice never knows {\it how} the qubit arrived at its final state. 
Indeed,
%for $\A_0=\A_1$,
even the probability of the most likely reading $f_2=\zz$ contains contributions from each path,
\begin{eqnarray}\label{g6}
\rr_{1\gets 0}(f_2=\zz)=\frac{|\A_0G_2(\zz)+\A_1G_2(\zz-1)|^2}{|\A_0+\A_1|^2}.
\end{eqnarray}
\par
However, Alice may gain information about a pre- and post-selected ensemble even without knowing the path chosen by the qubit. Having performed many trials (it will take more trials the larger is $\dd_2$), she can evaluate the average reading, i.e.,~first moment,
\begin{eqnarray}\label{g7}
\la f_2\ra_{1\gets 0} \equiv \int f_2 \tilde\rho_{1\gets 0} (f_2)df_2=\q\q\q\n
\frac{|\A_1|^2 +J_2\R[{\A_0}^*\A_1] }{|\A_0|^2+|\A_1|^2+ 2J_2\R[{\A_0}^*\A_1]}
{\xrightarrow[ \d_2 \to \infty ] {}}\zz.
\end{eqnarray}
{There is no} contradiction with the Uncertainty Principle, which permits knowing the amplitudes $\A_i$, 
[and, therefore, their particular combination (\ref{g3})]. What the principle forbids is using this knowledge
to answer, among~other things, the~\e{which way?} question. We illustrate this with the next~example. 

%%%%%%%%%%%%%%%%%%%%%%%%%%%%%%%%
\section{Two Inaccurate Quantum Pointers, and~Another Conclusion Not to~Make}\label{sec:9}
As in the classical case, Alice can employ at $t=t_2$ two highly inaccurate pointers $\dd_2 =\dd_5 \equiv \dd \gg 1$, which measure projectors on the states $|b_0\ra$ and $|b_1\ra$, 
respectively.
Now, by~Equation~(\ref{xx7}), the~distribution of the readings is Gaussian,% \vspace{-12pt}

\begin{eqnarray}\label{j1}
\rr_{1\gets 0}(f_2,f_5)=\q\q\q\q\q\q\q\q\q\q\n
\frac {|\A_0G_2(f_2)G_5(f_5-1)+\A_1G_2(f_2-1)G_5(f_5)|^2}{|\A_0|^2+|\A_1|^2+ 2J_2 J_5 
	\R\left [{\A}^*_0 \A_1\right ]}\n
{\xrightarrow[ \d \to \infty ] {}} 
G_2^2(f_2-\zz_2)G_5^2(f_5-\zz_5),\q\q
\end{eqnarray}
where
\begin{align}\label{j2}
\zz_2=\R\left [\frac{\A_1}{\A_0+\A_1}\right ], \q \zz_5=\R\left [\frac{\A_0}{\A_0+\A_1}\right ]=1-\zz_2.
\end{align}

%Consider, for simplicity, the case $\A_0=\A_1$. (This can be arranged, e.g.,~by choosing $|F_1\ra=|I_0\ra$, 
%$|b_{0,1}\ra=[|I_0\ra \pm |I_1\ra]/\sqrt{2}$, and putting $\u^s(t)=1$.)
%Now $z_2=z_5$ , and the most probable outcome is $(f_2=1/2, f_5=1/2)$. 

And, as~in the classical case, we encourage the reader to avoid the following reasoning (see Section~\ref{sec:8}):
%\begin{itemize}
\begin{enumerate}[label=(\roman*)]
	\item{A pointer \e{moves} (\e{weak trace} \cite{3P1} is produced) [cf. Equation~(\ref{g4})]
		only when the qubit is in the state upon which the projection is made. }
	% travels the path $\{ 1\gets 1\gets 1 \gets 0\}$. 
	% Similarly for the pointer with position $f_5$.}
	%\item{The most likely outcome $(f_2=1/2, f_5=1/2)$ suggests that both pointers \e{have moved} at the same time.}
	\item{Eq. (\ref{j1}) suggests that both pointers have moved (albeit on average).}
	\item{Hence, there is experimental evidence of the {qubit's presence in both states at $t=t_2$} and, therefore, {in both paths connecting $|I_0\ra$ with $|F_1\ra$.} }
	%\end{itemize}
\end{enumerate}
%The last statement is in contradiction with the Uncertainty Principle~\cite{FeynL}, \cite{FeynC}. The principle, we recall 
%forbids answering the \e{which way?} question, unless interference is destroyed. Here the perturbation produced by the inaccurate pointers 
%vanishes in the limit $\dd\to\infty$ yet Equation~(\ref{j1}) remains valid. 

{As in} the classical case, we find the fault with using the position of the maximum of the distribution (\ref{j1}).
As was shown in the previous section, an~inaccurate quantum pointer loses the \e{which way?} information. 
The information cannot, therefore, be recovered by employing two, or more, such pointers
to predict the presence of the qubit in a given~state. \par 
In~\cite{FeynC}, it was pointed out that assuming that in a double-slit experiment the particle passes through both slits at the same time
may lead to a wrong prediction. Namely, only a part of an electron, or~photon, would need to be detected at the exit of a slit, and~this is not what happens in practice.
Next, we briefly review the argument of~\cite{FeynC} in the present context. 
%Clearly, this cannot be done bearing in mind that the desired information was not present in each individual pointer's reading. 
\begin{figure}[H]
	\centering\includegraphics[angle=0,width=0.45\textwidth]{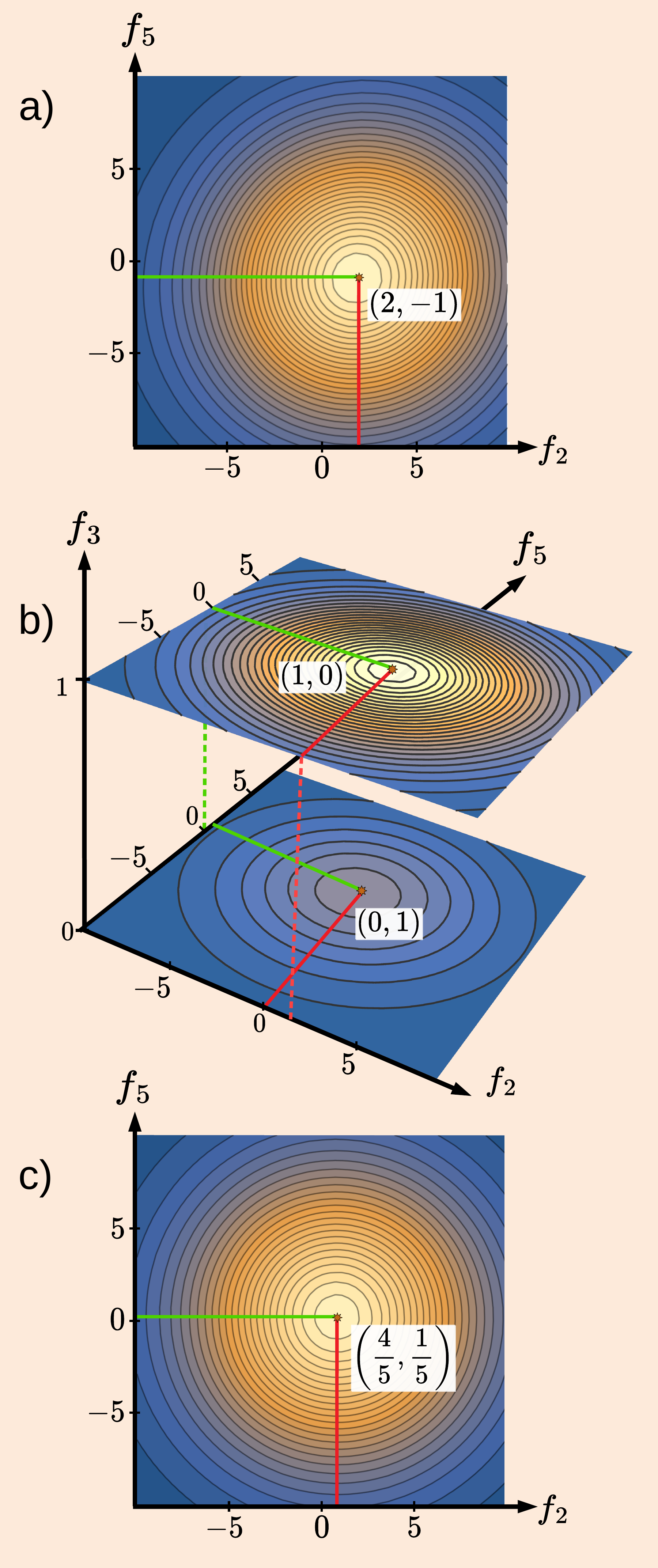}
	\caption{{(\textbf{a}) Distribution} 
		of the readings $\rr_{1\gets 0}(f_2,f_5)$ of two inaccurate quantum pointers with $\A_0=-\A_1/2$ and $\dd_2=\dd_5=10$,
		monitoring a qubit at $t=t_2$ [cf. Equation~(\ref{j1})]; (\textbf{b}) The distribution $\rr_{1\gets 0}(f_2,f_3,f_5)$ in Equation~(\ref{e1})
		with an accurate pointer, $\dd_3=0$, added at $t=t_3$. 
		(\textbf{c}) The result of integrating the distribution in ({b}) over $df_3$. Note that in ({c}), one does not recover the distribution shown in ({a}). 
	}
	\label{F5}
\end{figure}

%%%%%%%%%%%%%%%%%%%%%%%%%%%%%%%%%%%%%%%
\section{A \e{Wrong~Prediction}}\label{sec:10}
If not convinced by the argument of the previous section, Alice can follow the advice of~\cite{FeynC}, and~attempt to study the qubit's evolution in more detail.
In particular, she can add an accurate pointer acting at $t=t_3$ (see Figure~\ref{F4}), 
%to check whether in every trial it would 
in order to
detect only a part of the qubit travelling along the path $\{ F_1\gets c_1\gets b_1 \gets I_0\}$. If~the distribution (\ref{j1}) 
%shown in Figure~\ref{F5} 
is a proof of the qubit being 
%both in $|b_0\ra$ and $|b_1\ra$ 
present in both paths
at $t_2$
in any meaningful sense, this must be the only logical expectation. 
Since there is no path connecting $|b_0\ra$ with $|c_1\ra$ (see Figure~\ref{F4}),
two parts of the qubit cannot recombine in $|c_1\ra$ at $t_3>t_2$. 
% In Figure~\ref{F3} no path connects the states $|b_0\ra$ and $|c_1\ra$, and the qubit's parts passing through $|b_0\ra$ and $|b_1\ra$, 
%cannot be recombined to produce an entire qubit in $|c_1\ra$ at $t=t_3$. 
\par
However, at~$t=t_3$, Alice finds either a complete qubit, or~no qubit at all. As $\dd_3 \to 0$, the~distribution (\ref{c5}) becomes bimodal
in a three-dimensional space ($f_2$,$f_3$,$f_5$)%\vspace{-12pt}
\begin{align}\label{e1}
\rr_{1\gets 0}(f_2,f_3,f_5)
%\frac {|\A_0G_2(f_2)G_3(f_3)G_5(f_5-1)+\A_0G_2(f_2-1)G_5(f_5)|^2}{|\A_0|^2+|\A_1|^2+ 2J_2 J_3J_5 
%\R\left [{\A}^*_0 \A_1\right ]}\n
{\xrightarrow[ \d_3 \to 0 ] {}} 
\frac {|\A_0|^2G^2_2(f_2)\delta(f_3)G^2_5(f_5-1)}{|\A_0|^2
	+|\A_1|^2}&\\
+\frac {|\A_1|^2G^2_2(f_2-1)\delta(f_3-1)G^2_5(f_5)}{|\A_0|^2+|\A_1|^2}&.\nonumber
\end{align}
{Possible values} of $f_3$ are $0$ and $1$, and~only one of the pointers acting at $t=t_2$ is seen
to \e{move} in any given trial. 
Contour plots of the densities in Equations~(\ref{j1}) and (\ref{e1}) are shown in Figure~\ref{F5}(\textbf{a}),(\textbf{b}), respectively.
Note that integrating the density in Figure~\ref{F5}(\textbf{b}) over $df_3$ does not reproduce that in Figure~\ref{F5}(\textbf{a}) but~rather 
the density (\ref{f1}) [cf. Figure~\ref{F3}(\textbf{a})] for a classical system with $P_0=1/5$ and $P_1=4/5$, shown in Figure~\ref{F5}(\textbf{c}).

One can still argue that Alice does not compare like with like since the added accurate pointer perturbs the qubit's evolution
in a way that makes it choose the path $\{F_1\gets c_1\gets b_1 \gets I_0\}$. The~difficulty with this explanation 
is well known in the analysis of delayed choice experiments~\cite{DL1}. The decision to couple the accurate pointer may be taken by Alice after $t_2$
and cannot be expected to affect the manner in which the qubit passes through the states $|b_0\ra$ and $|b_1\ra$. 
This argument usually serves as a warning against a naive realistic picture for interpreting quantum phenomena~\cite{DL1}.
The conclusion in the case studied here is even simpler. \e{Weak traces} are not faithful indicators of the system's presence 
at a given location, and~using them as such leads to avoidable contradictions.

%There is still the approach of Sect.VI Alice can use in order to rationalise qubit's behaviour, as we will discuss next. 
%%%%%%%%%%%%%%%%%%%%%%%%%%%%%%%%%%%%%%%%
\section{Quantum \e{Hidden~Variables}}\label{sec:11}
In order to keep the rate of successful post-selections intact, $W_1(t_4)=|\A_0+\A_1|^2$, Alice may only use weakly coupled and, therefore, 
inaccurate pointers. These, as~was shown above, yield no information 
as to whether the qubit is in the state $|b_0\ra$ or $|b_1\ra$ at $t=t_2$ in any given trial, 
so the question remains unanswerable in principle.
The same is true for classical inaccurate pointers in Section~\ref{sec:6}, but~there it is possible to deduce the probabilities, $P_0$ and $P_1$, 
with which the system travels each of the two paths [cf. Equation~(\ref{cc0})].
In the quantum case, an~attempt to find directly unobservable \e{hidden} path probabilities governing the statistical behaviour of an unobserved 
system fails for a simple reason. 
With no a priori restrictions on the signs of $\A_i$, the~measured 
%\e{weak value} 
$\zz$ in Equation~(\ref{g3}) can have any real value (see~\cite{Sokolovski_2023_1}). 
For a negative $\zz$, 
the \e{probability} ascribed to the path $\{ F_1\gets c_1\gets b_1 \gets I_0\}$,
\begin{align}\label{s1}
P_1=\zz W(t_4) < 0
\end{align}
will also have to be negative. 
Thus, $P_1$ cannot be related to a number of cases in which the system follows the chosen path~\cite{FeynComp}, 
and a realistic explanation of the double-slit phenomenon fails as~expected. 
%%%%%%%%%%%%%%%%%%%%%%%%%
\section{Summary and~Discussion}\label{sec:12}
In summary, a~weakly coupled pointer employed to monitor a quantum system is, by necessity, an~inaccurate one. 
As such, it loses information about the path taken by the system in any particular trial,
yet one can learn something about path probability~amplitudes. 

A helpful illustration is offered by a classical case, where a stochastic two-way system is observed
by means of a pointer, designed to move only if the system takes a particular path,
leading to a chosen destination. 
The pointer can be rendered inaccurate by making its initial position random.
%, distributed around the zero setting.
%The final distribution of the readings is a weighted sum of the original distribution, and of its shifted copy. 
For an accurate pointer, the~final distribution of the reading consists of two non-overlapping parts, and~one always 
knows which path the system has travelled.
For a highly inaccurate pointer, the~final distribution is broad, and~not a single reading can be attributed 
to one path in preference to the other. 
\par
The distribution of the initial pointer's positions can be chosen to be a Gaussian centred at the origin.
It is a curious property of broad Gaussians that the final pointer's reading repeats the shape 
of the original distribution [cf. Equation~(\ref{h1a})], shifted by a distance equal to the probability of travelling the 
chosen path, conditioned on reaching the desired destination [cf. Equation~(\ref{h1b})]. 
The transition from two maxima to a single peak, achieved when the width of the Gaussian reaches the critical value, is sudden, 
and can be described as the cusp catastrophe [see Appendix \ref{appx:B}].
Thus, although~the \e{which way} information
is lost in every trial, one is still able to determine parameters (path probabilities) of the relevant statistical ensemble,
e.g., by~looking for the most probable final reading, or~by measuring the first moment of the distribution. 
For a broad Gaussian, these tasks would require a large number of trials. 
\par
The same property of the Gaussians may be responsible for a false impression that two inaccurate pointers [cf. Equations~(\ref{f1}) and Figure~\ref{F3}a]
move simultaneously (albeit on average), and~that this indicates the presence of the system in both paths at the same time. 
The fallacy is easily exposed 
by employing one more accurate pointer (see Figure~\ref{F3}b), or~simply by recalling that the system cannot be split in two. 
\par
Although the quantum case is different, parallels with the classical example can still be drawn.
%Classical observation can be non-invasive, while quantum measurements are not, and one needs to specify
% the interaction with a measuring device, as was done in Equations~(\ref{}), 
The accuracy of a quantum pointer depends 
on the uncertainty of its initial position, i.e.,~on the wave function (\ref{c2}). Weakening the coupling between the pointer and the system has the same effect as broadening the initial state. 
The distribution of the readings of an accurate pointer consists of two disjoint parts, and~one always knows which path has been taken, 
at the cost of altering the probability of a successful post-selection. The~only way to keep the probability intact is to reduce the coupling to (almost) zero,
but then there is not a single reading which can be attributed to a particular path. 
\par
Owing to the already mentioned property of Gaussians (see Appendix \ref{appx:D}), the~most likely reading of a highly inaccurate pointer is given by the real part 
of a quantum 
\e{weak value} (\ref{g3}), the~relative (i.e., normalised to a unit sum) path amplitude. 
%With no {\it apriori} restrictions on either magnitude or the sign of these amplitudes. 
Unlike the classical \e{weak values} in Equation~(\ref{h1c})
which must lie between $0$ and $1$, 
their quantum counterparts %in Equation~(\ref{g3})
can have values anywhere in the complex plane~\cite{Sokolovski_2023_1}.
As in the classical case, employing a weakly coupled quantum pointer 
%looses the \e{which way?} information, yet 
allows one to determine certain
parameters (probability amplitudes rather than probabilities) of the quantum ensemble [cf. Equations~(\ref{cc0})~and~(\ref{s1})]. 
\par
Equally inadvisable is using the joint statistics of two weak inaccurate quantum pointers [cf. Equation~(\ref{j1})] as~evidence of the quantum system's presence 
in both pathways at the same time, firstly for~reasons similar to those discussed in the classical case and secondly since this would lead to a wrong prediction~\cite{FeynC}. An~additional accurate pointer always detects either an entire qubit, or~no qubit at all, albeit at the price
of destroying interference between the paths. In~the setup shown in Figure~\ref{F4}, the~parts of the qubit, presumably present in both paths, 
%before the accurate measurement was made,
have no means to recombine by the time the accurate measurement is made, hence a contradiction. A~similar problem occurs with the interpretation of delayed choice experiments~\cite{DL1}, to~which we refer the interested reader. 
\par
Our concluding remarks can be condensed to few sentences.
Unlike the probabilities, the~probability amplitudes, used to describe a quantum system, are always available to a theorist. 
Weak measurements only determine the values of probability amplitudes, or~of their combinations. 
The Uncertainty Principle forbids one to determine the path taken by a quantum system, unless~interference between the paths is destroyed~\cite{FeynL}.
%, without destroying the interference
Hence, the weak values have little to contribute towards the resolution of the quantum \e{which way?} conundrum.

%%%%%%%%%%%%%%%%%%%%%%%%%%%%%%%%%%%%%%%%%%
%\section{Patents}
%
%This section is not mandatory, but may be added if there are patents resulting from the work reported in this manuscript.

%%%%%%%%%%%%%%%%%%%%%%%%%%%%%%%%%%%%%%%%%%
\vspace{6pt} 

%%%%%%%%%%%%%%%%%%%%%%%%%%%%%%%%%%%%%%%%%%
%% optional
%\supplementary{The following supporting information can be downloaded at:  \linksupplementary{s1}, Figure S1: title; Table S1: title; Video S1: title.}


% Only used for preprtints:
% \supplementary{The following supporting information can be downloaded at the website of this paper posted on \href{https://www.preprints.org/}{Preprints.org}.}

%\begin{tabularx}{\textwidth}{lll}
%\toprule
%\textbf{Name} & \textbf{Type} & \textbf{Description} \\
%\midrule
%S1 & Python script (.py) & Script of python source code used in XX \\
%S2 & Text (.txt) & Script of modelling code used to make Figure X \\
%S3 & Text (.txt) & Raw data from experiment X \\
%S4 & Video (.mp4) & Video demonstrating the hardware in use \\
%\bottomrule
%\end{tabularx}
%}

%%%%%%%%%%%%%%%%%%%%%%%%%%%%%%%%%%%%%%%%%%
%\authorcontributions{All authors contributed equally to this work.}
\bibliography{sn-bibliography.bib}

\section{Acknowledgenments}
{D.S. acknowledges financial support by the Grant PID2021-126273NB-I00 funded by MICINN/AEI/10.13039/501100011033 and by "{ERDF A way of making Europe}", as well as by the Basque Government Grant No. IT1470-22.\newline
	A.U. and E.A. acknowledge the financial support  by MICIU/AEI/10.13039/501100011033 and FEDER, UE through BCAM Severo Ochoa accreditation CEX2021-001142-S / MICIU/ AEI / 10.13039/501100011033; “PLAN COMPLEMENTARIO MATERIALES AVANZADOS 2022-2025 “, PROYECTO $ \text{N}^\text{o} $:1101288 and grant PID2022-136585NB-C22; as well as by the Basque Government through ELKARTEK program under Grants KK-2023/00017, KK-2024/00062 and the BERC 2022-2025 program. This work was also supported by the grant BCAM-IKUR, funded by the Basque Government by the IKUR Strategy and by the European Union NextGenerationEU/PRTR.}

\newpage
\appendix
\section{Some properties of Gaussian distributions }\label{appx:A}
Consider a function 
\begin{align}\label{x1}
F(f)=A\exp\left [-\frac{(f-a)^2}{(\d)^2}\right]+B\exp\left [-\frac{(f-b)^2}{(\d)^2}\right] \q\q
\end{align}
with arbitrary real $A$, $B$, $a$, and $b$. For $\d \ll |b-a|$, $F$ has two maxima at $f=a$ and $f=b$, 
and a single minimum between them. We are interested in the opposite limit, 
$\d \gg|b-a|$, where $F$ has a single maximum at 
\begin{align}\label{x2_1}
z = \frac{aA+bB}{A+B},
\end{align}
easily found by solving $\partial_f F(f)=0$ in the limit $\d\to \infty$.
Note that if  $A$ and $B$ have opposite signs, 
$z$ can lie outside the interval $[a,b]$.
In fact, $F(f)$ can be approximated by a single Gaussian 
\begin{align}\label{x2}
F(f){\xrightarrow[ \d \to \infty ] {}}\F(f)=(A+B)\exp\left [-\frac{(f-z)^2}{(\d)^2}\right],
\end{align}
to which it converges point-wise. 
Indeed, putting $x=f/\d$, 
and expanding the exponentials in Eqs.(\ref{x1}) and (\ref{x2}) 
in Tailor series we find
\begin{eqnarray}\label{x3}
\left  |\frac{F(x)-\F(x)|}{F(x)}\right | \approx \frac{(b-a)^2(2x^2-1)}{(\d)^2} \n
\times\left |\frac{AB}{(A+B)^2}\right | +o((\d)^{-2}),\q
\end{eqnarray}
so that the relative error of the approximation (\ref{x2}) can be made small 
for any given $f$.  
\newline
We note further that in the limit  $\d \to \infty $, the first moments $F$ and $\F$ agree
[$\la f^n\ra_F \equiv \int f^n F(f)df/\int F(f)df$, $ n=1,2,... $]
\begin{align}
\la f\ra_{F}= \la f\ra_{\F}=z,
\end{align}
\newline
but the second moments, each of order of $(\d)^2$, differ by a finite quantity, 
\begin{align}\label{x5}
\la f^2\ra_F - \la f^2\ra_{\F}=\frac{AB(a-b)^2}{A+B},
\end{align}
so $F$ and $\F$ can, at least in principle, be distinguished. \newline
\newline
An approximation, similar to the one in Eq.(\ref{x2}), can also be obtained in two dimensions, 
by considering 
\begin{align}\label{x6}
F(\f)=A\exp\left [-\frac{(\f-\aa)^2}{(\d)^2}\right]+B\exp\left [-\frac{(\f-\bb)^2}{(\d)^2}\right],
\end{align}
where $\vec y=(y_1,y_2)$ is a two dimensions vector, and $(\vec y)^2\equiv y_1^2+y_2^2$.
As $\d \to \infty$ we find 
\begin{align}\label{x7}
F(\f){\xrightarrow[ \d \to \infty ] {}} (A+B) \exp\left [-\frac{(\f-\vec z)^2}{(\d)^2}\right]
\end{align}
where 
\begin{align}\label{x8}
z_i=\frac{a_iA+b_iB}{A+B}, \q i=1,2.
\end{align}
%%%%%%%%%%%%%%%%%%%%%%%
\section{Connection with catastrophe theory}\label{appx:B}
\begin{figure}[h!]
	\centering\includegraphics[angle=0,width=.46\textwidth]{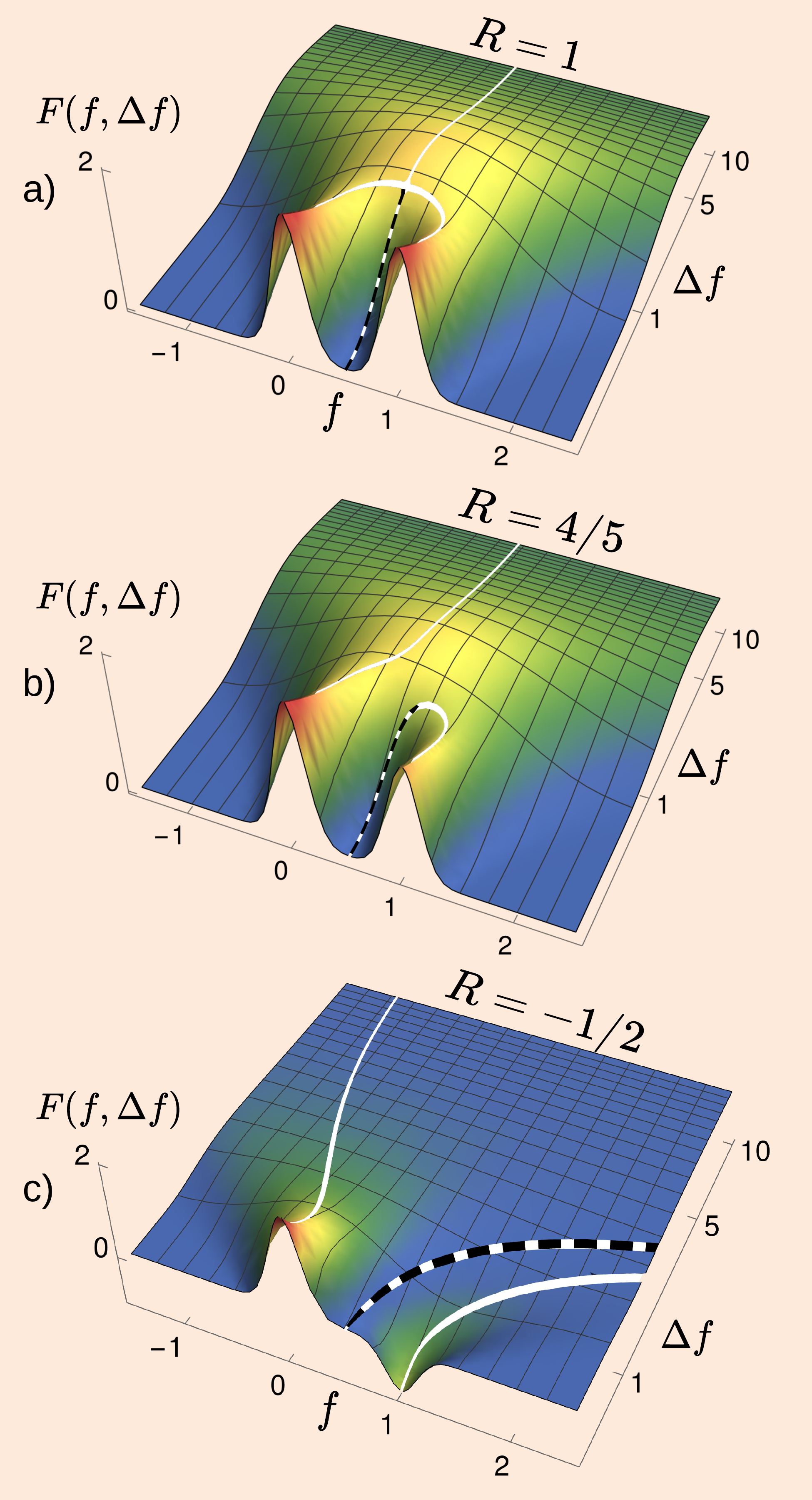}
	\caption{Maxima (solid white) and minima (dashed) of $F(f)$ vs. $f$ and $\d$ (a.u.).
		(\textbf{a}) Pitchfork bifurcation for $R=1$, (\textbf{b}) same as (\textbf{a}) but for $R=4/5$. (\textbf{c}) same as ({b}) but for a negative $R=-1/2$.
		Note that in this case one has an \e{anomalous} value $z=-1$.}
	\label{F6}
\end{figure}

%%%%%%%%%%%%%%%%%%%%%%%%%%%%%%%%%%%%%
%%%%%%%%%   TEXT    %%%%%%%%%
To study  the transformation of two maxima and a minimum of the function $F(f)$ in Eq.(\ref{x1}) into a single maximum, we choose a special case $a=0$ and $b=1$.
The structure of the extrema of $F(f)$ is  determined by two parameters, $R=B/A$ and $\Delta f$, and corresponds, therefore,  to cusp singularity case of the Catastrophe Theory \cite{Cata}. In the symmetric case, $R=1$, $F(f)=F(f+1)$ and there is always a single extremum at $f=1/2$. The first and second derivatives at  $f=1/2$ are given by 
\begin{align}\label{xx1}
\partial _f F\left (\frac{1}{2},\d\right)&=0,\\
\partial^2 _f F\left (\frac{1}{2},\d\right)&=-\frac{4e^{-\frac{1}{4(\d)^2}}}{(\Delta f)^{4}}\left((\Delta f)^2-\frac{1}{2}\right),\nonumber
\end{align}
and the three extrema coalesce at $\Delta f = \sqrt {\frac{1}{2}}$, where $\partial _f F(1/2,\d)=0$.
This is a case of pitchfork bifurcation \cite{Cata}, shown in Fig.\ref{F6}(\textbf{a}). Other cases are shown in Figs.\ref{F6}(\textbf{b}) and \ref{F6}(\textbf{c}).
Note that with $R<0$ the single maximum which survives as $\Delta f \to \infty$ lies outside the interval $0 \le f\le 1$.
%%%%%%%%%%%%%%%%%%%%%%%%
\section{The likelihood of discovering which way a classical system went }\label{appx:C}
Consider a probability distribution,  
\begin{align}\label{B1}
\rho(f)=\frac{1}{2(\d) \sqrt \pi}\left \{\exp\left [-\frac{f^2}{(\d)^2}\right]+\exp\left [-\frac{(f-c)^2}{(\d)^2}\right] \right \}, \q\q
\end{align}
with $c>0$, and search for a range of $f$, where the first term 
can be neglected, compared to the second one. 
For example, one may look for readings where 
the ratio  between the two terms
does not exceed some  $\ep \ll1$. Such readings would occur for  $f > f_\ep$, where
% \begin{eqnarray}\label{B2}
%$\exp\left [-\frac{(2f_\ep c-c^2)}{\d^2}\right]=\ep$, or
%\end{eqnarray}
%\begin{eqnarray}\label{B2}
$ f_\ep=(\d)^2|\ln \ep|/2c +c/2.$ 
%\end{eqnarray}
%Thus the likelihood of finding a reading whose probability 
%is determined only by the second exponential  in (\ref{B1}) is
The probability of finding a reading of this kind is, therefore, 
$P(f>f_\ep) =\int_{f_\ep}^\infty\rho(f)df$. Replacing  
$\exp\left [-\frac{f^2}{(\d)^2}\right]$ by a larger term $\exp\left [-\frac{(f-c)^2}{(\d)^2}\right]$, 
we have [$\text{erfc}(x)$ is the complementary error function \cite{ABRAM}]
\begin{align}\label{B2}
P(f>f_\ep)< 2^{-1}\text{erfc}\left[\frac{f_\ep-c}{(\d)}\right ]{\xrightarrow[ \d \to \infty ] {}}&\\ 
\frac{c}{2\sqrt \pi (\d)|\ln \ep|}\exp\left [-\frac{(\d)^2|\ln \ep|^2}{4c^2}\right]\to 0&.\nonumber
\end{align}
The probability of finding a value $f$, which can be attributed to {only
	one} of the two terms in Eq.(\ref{B1}), vanishes  rapidly for $(\Delta f)\gg c$. 
%\section{Classical normalisation of pre-selected ensembles}\label{Appx:normalisation}
%\begin{align}
%&\rho(f_1,f_2,f_3,f_4)=\nonumber\\
%&\sum_{i,j,k,l}G_1(f_1-i)H(f_2-j)J(f_3-k)K(f_4-l)G_5(f_5-1)\nonumber\\
%&P(l\gets k\gets j \gets i)w_i.
%\end{align}
%where $w_i P(l\gets k\gets j \gets i)=w_i P(i\to j)P(j\to k)P(k\to l) $.
%
%\begin{align}
%\tilde{\rho}(1,f_2,f_3,f_4)=\sum_{j,k,l}G(f_1^1)H(f_2-j)J(f_3-k)K(f_4-l)P(l\gets k\gets j \gets 1)
%\end{align}
%where the sum over $ i $ is no longer a sum\footnote{Note that the sum over $ i $ disappears as a consequence of having an strong measurment, 
%	\begin{align}
%	\sum_i G(f_1-i)\to\sum_i G(f^1_1-i)=G(1-0)+G(1-1)=G(f_1^1-1)
%	\end{align} } and is unnormalised (since we forced one of the initial states, i.e. $ f_1=1 $), i.e. $ \int df_{1,2,3,4}\tilde{W}<\int df_{1,2,3,4}{W} $, let us normalize it:
%\begin{align}
%&{\rho}(1,f_2,f_3,f_4)=\frac{\tilde{W}}{\iiint \tilde{W} df_2df_3df_4}\\
%&=\frac{\sum_{j,k,l}G(f_1^1-1)H(f_2-j)J(f_3-k)K(f_4-l)P(l\gets k\gets j \gets 1)}{\sum_{j,k,l}P(l\gets k\gets j \gets 1)}\\
%&=\frac{{G(f_1^1-1)}\sum_{j,k,l}H(f_2-j)J(f_3-k)K(f_4-l) {w_1} P(1\to j)P(j\to k)P(k\to l)}{ G(f_1^1-1)\sum_{j,k,l}{w_1} P(1\to j)P(j\to k)P(k\to l)}\\
%&={\sum_{j,k,l}H(f_2-j)J(f_3-k)K(f_4-l) P(1\to j)P(j\to k)P(k\to l)}
%\end{align}
%%%%%%%%%%%%%%%%%%%
\vspace{.1cm}
\section{More properties of Gaussian distributions }\label{appx:D}
Consider next a function
\begin{align}\label{xx1}
F(f)=\left |A\exp\left [-\frac{(f-a)^2}{(\d)^2}\right]+B\exp\left [-\frac{(f-b)^2}{(\d)^2}\right] \right |^2
\end{align}
with complex valued $A$ and $B$, and real $a$ and $b$
\begin{align}\label{xx2}
A=A_R+iA_I, \q B=B_R+iB_I.
\end{align}
Equation (\ref{xx1}) can be rewritten as
\begin{align}\label{xx3}
F(f)=\left \{A_R\exp\left [-\frac{(f-a)^2}{(\d)^2}\right]+B_R\exp\left [-\frac{(f-b)^2}{(\d)^2}\right] \right \}^2\nonumber\\
+\left \{A_I\exp\left [-\frac{(f-a)^2}{(\d)^2}\right]+B_I\exp\left [-\frac{(f-b)^2}{(\d)^2}\right] \right \}^2.
\end{align}
Applying (\ref{x2}) to each  term in the curly brackets, and then to the sum of the results, yields
\begin{align}\label{xx4}
F(f) {\xrightarrow[ \d \to \infty ] {}}\tilde{F}(f)= |A+B|^2\exp\left [-2\frac{(f-\R[z])^2}{(\d)^2}\right],
\end{align}
with $z$ still given by Eq.(\ref{x2_1}), but with complex valued $A$ and $B$, 
\begin{align}\label{xx5}
z= \frac{a(A_R+iA_I) +b(B_R+iB_I)}{(A_R+B_R) +i(A_I+B_I)}. 
\end{align}
%{\r
%	We also know that it peaks in $z$ calculating the $\tilde f$ fulfilling.
%	\begin{align}
%	\eval{\frac{d\left(F^*F\right)}{df}}_{\tilde{f}}&=0,\\
%	\tilde{f}{\xrightarrow[ \d \to \infty ] {}}&\Re\left(\frac{Aa+Bb}{A+B}\right),
%	\end{align}
%where $ \rho=F^*F $ as $ F $ 
Extension to two dimensions can be done in a similar manner as in Appendix \ref{appx:A}. For
$\f=(f_1,f_2)$, $\aa=(a_1,a_2)$ and $\bb=(b_1,b_2)$
\begin{align}\label{xx6}
F(\f)=\left |A\exp\left [-\frac{(\f-\aa)^2}{(\d)^2}\right]+B\exp\left [-\frac{(\f-\bb)^2}{(\d)^2}\right] \right |^2.
\q
\end{align}
we find
\begin{align}\label{xx7}
F(\f) {\xrightarrow[ \d \to \infty ] {}} |A+B|^2\exp\left [-2\frac{(\f-\R[\vec z])^2}{(\d)^2}\right],
\end{align}
where $\vec z$ is still given by Eq.(\ref{x8}) with complex valued $A$ and $B$.

\end{document}